\documentstyle[epsfig]{disk}
\begin{document}

\title{Disc instability models, evaporation and radius variations}

\author{Jean-Marie HAMEURY \\ 
{\it Observatoire de Strasbourg, 11 rue de l'Universit\'e, F-67000 Strasbourg,
France, hameury@astro.u-strasbg.fr} \\
Guillaume DUBUS \\
{\it Anton Pannekoek Astronomical Institute, 
University of Amsterdam, Kruislaan 403,
1098 SJ Amsterdam, Netherlands, gd@astro.uva.nl}
Jean-Pierre LASOTA \\
{\it D\'epartement d'Astrophysique Relativiste et de Cosmologie, Observatoire de
Paris, F-92195 Meudon C\'edex, France, lasota@obspm.fr}
Kristen MENOU \\
{\it Harvard-Smithsonian  Center for Astrophysics, 60 Garden Street,
Cambridge, MA 02138, USA, kmenou@cfa.harvard.edu}}

\maketitle

\section*{Abstract}

We show that the outcome of disc instability models is strongly influenced by
boundary conditions such as the position of the inner and outer disc edges. 
We discuss other sources of uncertainties, such as the tidal torque, and we
conclude that disc illumination, disk size variations and a proper
prescription for the tidal torque must be included in models if one wishes to
extract meaningful physical information on e.g. viscosity from the comparison
of predicted and observed lightcurves.

\section{Introduction}

The bases of the thermal-viscous accretion disc instability have been
established 25 years ago (see Smak in these proceedings for an historical
overview), and despite the successes of this model in explaining dwarf nova
(DN) outbursts and, to a lesser extent, soft X-ray transients (SXTs), there
are still a number of observational facts that conflict the predictions of
the standard version of the model. One can mention, for example, the
detection of a very significant quiescent X-ray flux in both DNs and SXTs at
a level that exceeds theoretical predictions by up to six orders of magnitude
(Lasota, 1996). One could also quote the existence of steady bright X-ray
sources which should be unstable, or of ER UMa systems in which the
supercycle appears to be too short to be accounted for by the tidal-thermal
instability model. Finally, one should mention the fact with physically
plausible parameters most authors produce light curves (often not published)
that bear no resemblance whatsoever to any observed one.

Some of these discrepancies have already received an explanation : the X-ray
detection of systems in quiescence requires the inner disc radius be larger
than that of the compact object as a result of either evaporation or of the
presence of a magnetosphere; the stability of the accretion disc in bright
X-ray sources is influenced by illumination from the central X-ray source.
Other remain, such as the difficulty of reproducing similar peak luminosities
in systems exhibiting various outbursts shapes. One should also keep in mind
that, despite recent progress, the very nature of viscosity in accretion
discs remains largely unknown and that the $\alpha$ prescription is a mere
parameterization of our ignorance. It is thus extremely important to
disentangle numerical and physical effects when modeling accretion discs if
one wishes to infer a refined viscosity prescription from observations of
accretion disc outbursts.

In this paper, we describe our code, giving its advantages and limitations;
we then discuss the influence of physical effects such as variations of the
inner and outer disc radius. We show that all these effects (adding
irradiation) must be accounted for; these might play a major role in
superoutburtsts of SU UMa and ER UMa systems.

\section{Numerical modeling}
\subsection{Basics}

The numerical code used here is described in detail in Hameury et al. (1998),
and we recall briefly here its main characteristics. We solve the standard
mass continuity and angular momentum conservation equation of a
geometrically thin Keplerian disc, in which we take into account the tidal
torque, assumed to be of the form (Smak 1984; Papaloizou \& Pringle 1977):
\begin{equation}
T_{\rm tidal} = c \omega r \nu \Sigma \left( {r \over a} \right)^5
\end{equation}
where $\omega$ is the angular velocity of the binary motion, $r$ the distance
to the compact object, $\nu$ the viscosity, $\Sigma$ the surface density in
the disc, and $a$ the orbital separation. The constant $c$ is a free
parameter that is taken such as to give the required average outer disc
radius. The thermal equation is taken from
Cannizzo (1993):
\begin{equation}
{\partial T_{\rm c} \over \partial t} = { 2 (Q^ + -Q^- + J) \over C_P \Sigma}
 - {\Re T_{\rm c}
\over \mu C_P} {1 \over r} {\partial (r v_{\rm r}) \over \partial r} -
v_{\rm r} {\partial T_{\rm c} \over \partial r},
\end{equation}
where $T_{\rm c}$ is the central disc temperature, $Q^+$ and $Q^-$ are the
surface heating and cooling rates respectively, $C_P$ is the heat capacity at
constant pressure, $v_{\rm r}$ is the radial velocity and $J$ is a radial
energy flux which we assume here to be carried by turbulent eddies, and which
takes the form:
\begin{equation}
J = 1/r \partial / \partial r (r {3 \over 2} \nu C_P \Sigma {\partial 
T_{\rm c} \over \partial r}).
\end{equation}

The boundary conditions for the energy equation are unimportant as this
equation is dominated by the $Q^+$ and $Q^-$ terms except in the transition
fronts; we take $\partial T / \partial r = 0$ to minimize numerical problems.
We have as usual $\Sigma = 0$ at the inner edge, and we assume that matter is
added at the very outer edge, whose position $r_{\rm out}$ is not specified,
which gives two conditions at the outer edge:
\begin{equation}
\dot{M}_{\rm tr} = 2 \pi r \Sigma (\dot{r}_{\rm out} - v_{\rm})
\label{eq:bc1}
\end{equation}
and
\begin{equation}
\dot{M}_{\rm tr} \left[ 1 - \left( {r_{\rm k} \over r_{\rm out}}\right)^{1/2}
\right] = 3 \pi \nu \Sigma,
\label{eq:bc2}
\end{equation}

These equations with their boundary conditions are solved using a fully
implicit code in which equations are discretized on an adaptative grid
defined such as to resolve the temperature and density gradients. As a
result, heating and cooling fronts are always resolved. Moreover, as the code
is implicit, large time steps can be used, whatever the spatial resolution,
contrary to explicit codes for which the time step has to be smaller than the
thermal time ($1/\alpha$ times the dynamical time) at the inner edge. This
allows us to cover a large number of cycles, so that the memory of the initial
conditions has been lost.

The results obtained using this code do reproduce a number of dwarf nova
outburst characteristics, such as the occurrence of inside-out outbursts for
low mass transfer rate, and outside-in outbursts for larger values of
$\dot{M}_{\rm tr}$. However, we also obtain small, inside-out, intermediate
outbursts for high primary masses (i.e. low inner disc radius). Such weak
outburst are not observed, which means that we might be missing an
essential ingredient when modeling these outbursts.

\subsection{Heating and cooling fronts}

Because we use an adaptative grid, we are able to resolve both the heating
and cooling fronts. We confirm (Menou, Hameury \& Stehle 1998) that the
propagation velocity of these fronts is of order of $\alpha c_s$, where $c_s$
is the sound speed in the hot gas, with heating fronts propagating more
rapidly than cooling fronts. Moreover, whereas the general properties of
cooling fronts do not vary from one outburst to the other, those of heating
fronts depend sensitively upon the actual profile of $\Sigma$ as a function
of $r$.

We find that the width $w$ of the heating and cooling fronts is proportional
to the disc scale height $h$, contrary to Cannizzo, Chen \& Livio (1995) who
obtain $w \sim (hr)^{1/2}$; we get the same order of magnitude, but a
different radius dependence. $w$ is a few times $h$ for the heating fronts,
and much larger for cooling fronts, so that the thin disc equations still
apply.

We also confirm the self-similarity of the inner, hot disc during the cooling
phase that was found by Vishniac (1997), but we describe this disc property
in a slightly differentl way. $\Sigma$ is found to scale naturally
with $\Sigma_{\rm min}$, the minimum value of $\Sigma$ on the hot upper
branch in the $\Sigma$ -- $T_{\rm eff}$ diagram. In this regime, as already
noted by Vishniac (1997), the inner hot disc empties essentially by
transferring matter to the outer disc that has returned to the cold state,
and not by accretion onto the compact object. As a consequence, there is a
density jump in the passage of a cooling front; in the self-similar regime,
we find that $\Delta \log (\Sigma)$ is a constant. If this constant is larger
than $\log (\Sigma_{\rm max}/\Sigma_{\rm min})$, where $\Sigma_{\rm max}$ is
the maximum density on the cool branch, self-similarity can never be reached,
and reflares are obtained. This happens for high primary masses (a few
M$_\odot$) typical of black hole X-ray transients, as well as in cases where
the compact object is so hot that the inner portion is stabilized as a result
of illumination, as proposed by King (1997). In this case, $\Sigma_{\rm max}
= \Sigma_{\rm min}$ at the transition point between the hot inner disc and
the outer disc that can be subject to the thermal-viscous instability, and
many reflares are observed (Hameury, Lasota \& Dubus, 1999).

\section{Variations of the inner disc radius : evaporation, magnetospheres}

Both dwarf novae and soft X-ray transients emit a significant flux in X-rays
during quiescence (typically 10$^{32}$ erg~s$^{-1}$ for SXTs, and more than
10$^{30}$ erg~s$^{-1}$ for dwarf novae). As the whole disc must stay on the
cool branch during quiescence, the local mass transfer rate $\dot{M}$ must be
smaller than the critical value $\dot{M}_{\rm crit}$ above which this stable
cool branch disappears:
\begin{equation}
\dot{M} < \dot{M}_{\rm crit} = 4 \times 10^{15} \alpha_{\rm c}^{-0.04}
M_1^{-0.89} r_{10}^{2.67} \; \rm g \; s^{-1}
\end{equation}
where $\alpha_{\rm c}$ is the viscosity parameter on the cool branch, $M_1$
is the primary mass and $r_{10}$ is the radius in 10$^{10}$ cm. This can be
written as a constraint on the inner radius $r_{\rm in}$:
\begin{equation}
r_{\rm in} > 8 \times 10^8 \left({L_{\rm X} \over 10^{32} \; \rm erg \;
s^{-1}}\right)^{0.38} \left ({M_1 \over \rm M_\odot}\right)^{0.33} \left(
{\eta \over 0.1}\right)^{-0.38} \; \rm cm
\end{equation}
where $\eta$ is the efficiency of accretion; this is much larger than the
neutron star radius or the radius of the innermost stable orbit for SXTs, and
is also usually larger that the radius of the white dwarf in DNs.

Several mechanisms may cause the inner disc to be truncated; if the compact
object is a magnetized white dwarf, the Alfv\'en radius exceeds that of the
white dwarf in quiescence as soon as the magnetic moment of the white dwarf
is larger than 10$^{30}$ Gcm$^3$. Evaporation (see e.g. Meyer \&
Meyer-Hofmeister 1994; Liu, Meyer \& Meyer-Hofmeister 1992) leading to winds
or to the formation of an ADAF is also an attractive possibility, but
evaporation rates are quite uncertain.

These mechanisms can be accounted for by either introducing a mass loss rate
in the mass conservation equation, or by assuming that the inner disc radius
is a specified function of the mass accretion rate onto the compact object;
both prescriptions lead to qualitatively similar results. This differs a lot
from a situation in which a hot, geometrically thin and optically thick disc
would form close to the compact object as a result of e.g.  illumination
(King, 1997).

\subsection{The case of WZ Sge}

WZ Sge is a dwarf nova which exhibits long outbursts with a recurrence time
of about 30 years. Such a long recurrence time, together with the fact that
the amount of mass transferred during an outburst is large, requires a very
small value of the viscosity parameter $\alpha$ (typically $\alpha \sim
10^{-5} - 10^{-3}$) in the framework of the standard model (Smak, 1993;
Osaki, 1995; Meyer-Hofmeister, Meyer \& Liu, 1998), but the reason for such a
low $\alpha$ in this particular system is left unexplained.

Hameury, Lasota \& Hur\'e (1997a) proposed an alternative possibility with a
``normal" value of $\alpha$ ($\alpha = 0.01$) in which the inner part of the
accretion disc is disrupted by either a magnetic field (Lasota, Kuulkers \&
Charles 1999) or evaporation, so that the disc is stable (or very close to
being stable) in quiescence, as the mass transfer rate is very low and the
disc can sit on the cool lower branch of the thermal equilibrium curve. 
Outbursts are triggered by an enhanced mass transfer which renders the disc
unstable. The resulting outburst is strongly affected by the irradiation of
the secondary star, as the effective temperature of the irradiated hemisphere
is {\em observed} to reach 16,000 -- 17,000 K at maximum, i.e. increases by a
factor 10 with respect to quiescence. This results in a large increase of the
mass transfer rate; a straightforward application of the formulae given in
Hameury, King \& Lasota (1986) shows that $\dot{M}$ can increase by 2 to 3
orders of magnitude, and be comparable to the mass accretion rate onto the
white dwarf.

Numerical models in which one includes both evaporation (required by the long
recurrence time) and illumination (required to account for the total amount
of mass transferred) reproduce the overall light curve of WZ Sge.

\subsection{The case of GRO J1655-40}

GRO J1655-40 is a soft X-ray transient in which a 6 days delay between the
rise to outburst of optical and X-rays was observed (Orosz et al. 1997).
Hameury et al. (1997b) showed that the quiescent optical and X-rays
observations of this source indicate the presence of an ADAF which extends to
a radius of about 10$^{10}$ cm, with a mass transfer through the ADAF of a
few times 10$^{16}$ g~s$^{-1}$.

Hameury et al. (1997b) also show that the presence of such an ADAF that
disrupts the standard accretion disc in the vicinity of the black hole is
also required to account for the X-ray delay and the optical rise of the
outburst. If the disc were extending to the last stable orbit, one would
expect inside-out outbursts that de not produce any X-ray delay at all. If on
the other hand one assumes that the $\Sigma(r)$ density profile in the disc
is for some unspecified reason far from being relaxed, one could imagine that
large amounts of matter accumulate at the outer edge, and that outside-in
outbursts can occur. One can then reproduce the observed X-ray delay, but the
disc brightens much too fast in optical.

\subsection{The case of SS Cyg}

Whereas the interpretation of UV delays relies on a precise modeling of the
spectra of accretion discs, and hence on the way viscous heating is
distributed vertically in the disc, EUV being emitted by the boundary layer
can be directly translated into a mass accretion rate onto the compact
object. EUVE observations of SS Cyg by Mauche (1996) showed that there can be
a delay as large as 1 day between the EUV and the optical. Such a delay can
be easily accounted for if the inner parts of the disc evaporate at a
rate $\dot{M}_{\rm ev} = 1.5 \times 10^{16} (r/10^9 \; {\rm cm})^{-2}$ g
s$^{-1}$ (Hameury et al., 1999), in contrast with what would happen if the
sole effect of irradiation of the disc by the white dwarf were to heat it up.
In the latter case, the EUV delay is indeed longer than in the standard case,
but cannot be as long as one day; moreover, as mentioned earlier, reflares
are inevitable.

\section{Outer radius variations}

Hameury et al. (1998) have shown that numerical codes in which the outer
radius is allowed to vary produce results which are qualitatively different
from those in which $r_{\rm out}$ is kept fixed. When $r_{\rm out}$ is kept
fixed, one frequently obtains a large number of small outbursts
between major ones; in addition, it is extremely difficult to get outside-in
outbursts even for large mass transfer rates. The basic reasons for these
differences are (i) the fact that during an outburst, matter is pushed at
larger radius, implying a decrease of $\Sigma$ at the outer edge of the disc,
so that the cooling wave starts earlier, and the disc is less depleted, and
(2) the contraction of the disc in quiescence under the effect of the tidal
torque, which causes an increase of the local mass transfer rate to values
that can exceed the mass loss rate from the secondary. Both effects
facilitate the accumulation of matter in the outer region of the disc, and
hence lead to outside-in outbursts.

Such disc variations are observed during outburst cycles in dwarf novae, with
a rapid rise during outburst and a slow contraction of $\sim$ 20\%
during decline. However, radius variations are given by the tidal torque
$T_{\rm tidal}(r)$ for which no realistic analytic expression is available at
present, in particular close to the 3:1 resonance radius. This uncertainty is
particularly worrying for SU UMa systems, as $T_{\rm tidal}(r)$ is an
essential ingredient of the tidal-thermal instability (Osaki 1996).

\section{Conclusion}

\begin{figure}
\centerline{\psfig{figure=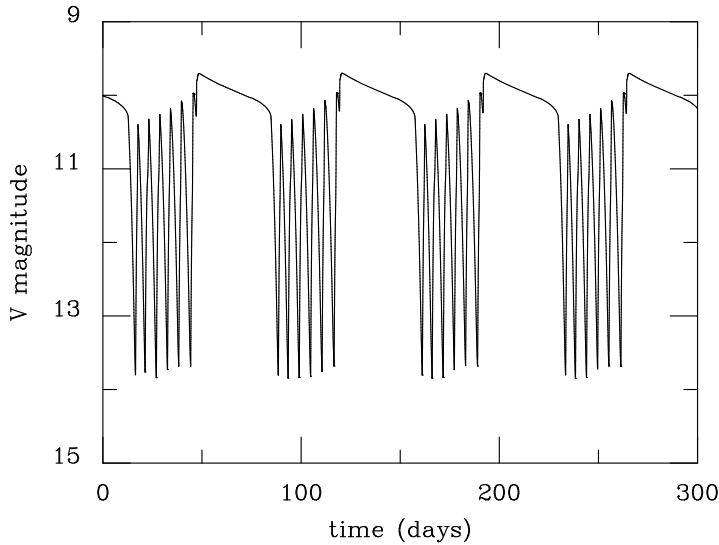,width=0.65\textwidth}}
\caption{Outbursts in a system with $M_1$ = 1 M$_\odot$, $\alpha$ = 0.02 on
the cool branch and 0.2 on the hot one, $r_{\rm in}$ = 5000 km, average $r_{\rm
out} = 1.3 \times 10^{10}$ cm, and illumination from the white dwarf with
temperature 35,000 K plus accretion energy. Secondary illumination has also 
been included.}
\end{figure}

Disruption of the inner parts of the disc by either evaporation or by a
magnetic field can easily account for the observed X-ray/EUV delays.  Such
holes in a disc are also required by X-ray observations of quiescent systems,
and by the long term cycles that do not exhibit small inside-out outbursts
that naturally appear when the inner radius of the disc is small. It also
very important to let the outer radius of the disc vary, and one would
require a better estimate of the tidal torque than presently available.
Finally, illumination by both the compact object and by the disc itself are
important, even in the case of cataclysmic variables. We show in Fig. 1 an
example of light curves in which all these effects have been included; as
can be seen, this curve is quite reminiscent of ER UMA systems, and has been
obtained without assuming any tidal instability. This does not mean that the
tidal-thermal instability model for superoutbursts is incorrect; it does
imply that effects that have been neglected until now must be added in models.

\section{References}
\re
Cannizzo J.K. 1993, ApJ 419, 318
\re
Cannizzo J.K., Chen W., Livio M. 1995, ApJ 454, 880 
\re
Hameury J.-M., Lasota J.-P., Hur\'e J.-M. 1997a, MNRAS 287, 937
\re
Hameury J.-M., Lasota J.-P., McClintock J.E., Narayan R. 1997b, ApJ 489, 234
\re
Hameury J.-M., Lasota J.-P., Dubus G. 1999, MNRAS, in press
\re
Hameury J.-M., Menou K., Dubus G., Lasota J.-P., Hure J.-M. 1998, MNRAS 298,
1048
\re
King A.R. 1997, MNRAS 288, L16
\re
Lasota J.P. 1996, in Proceedings of IAU Symposium 165, Compact Stars in
Binaries, J. van Paradijs et al (eds.), p. 43
\re
Lasota J.P., Kuulkers E., Charles P.A. 1999, MNRAS, submitted
\re
Liu B.F., Meyer F., Meyer-Hofmeister E. 1998, A\&A, submitted
\re
Mauche C. W. 1996, in Astrophysics in Extreme Ultraviolet, IAU Coll. 152, 
Bowyer S., Bowyer R.F. eds, Kluwer, Dordrecht, p. 317
\re
Menou K., Hameury J.-M., Stehle R. 1998, MNRAS, in press
\re
Meyer F., Meyer-Hofmeister E. 1994, A\&A 288, 175
\re
Meyer-Hofmeister E., Meyer F., Liu B. 1998, A\&A, in press
\re
Orosz J.A., Remillard R.A., Bailyn C.D., McClintock J.E. 1997, ApJ 478, L83
\re
Osaki Y. 1995, PASJ 47, 47
\re
Osaki Y. 1996, PASP 108, 39
\re
Papaloizou J., Pringle J.E. 1977, MNRAS 181, 441
\re
Smak J. 1984, Acta Astr. 34, 161
\re
Smak J. 1993, Acta Astr. 43, 101
\re
Vishniac E.T. 1997, ApJ 482, 414
\end{document}